\documentclass[a4paper,12pt]{article}

\usepackage{graphicx, amsmath, amsfonts, amssymb, array, natbib, graphicx, float, lipsum,tikz , lscape}

\usetikzlibrary{arrows,%
                shapes,positioning}
                
\title{Estimating the impact of treatment compliance over time on smoking cessation using data from ecological momentary assessments (EMA)}
\author{Yaoyuan V. Tan, Donna L. Coffman, Megan Piper, Jason Roy}

\begin{document}
\maketitle
%\linespread{2}
%\selectfont

%\begin{abstract}
%The Wisconsin Smoker’s Health Study (WSHS2) was a longitudinal trial conducted to compare the effectiveness of two commonly used smoking cessation treatments, varenicline and combination nicotine replacement therapy (cNRT) with the less intense standard of care, nicotine patch. The main outcome of the WSHS2 study was that all three treatments had equivalent treatment effects. However, in-depth analysis of the compliance data collected via ecological momentary assessment (EMA) were not analyzed. Compliance to the treatment regimens may represent a confounder as varenicline and cNRT are more intense treatments and would likely have larger treatment effects if all subjects  complied. In order to estimate the causal compliance effect, we view the counterfactual, the outcome that would have been observed if the subject was allocated to the treatment counter to the fact, as a missing data problem and proceed to impute the counterfactual. Our contribution to the methodological literature lies in the extension of this idea to a more general analytic approach that includes mediators and confounders of the mediator-outcome relationship. Simulation results suggest that our method works well and application to the WSHS2 data suggest that the treatment effect of nicotine patch, varenicline, and cNRT are equivalent after accounting for differences in treatment compliance. 
%\end{abstract}

\section{Introduction}
Smoking has been linked to many diseases including lung cancer and cardiovascular disease and is the leading cause of preventable death in the United States \citep{USDHHS}. Since the landmark US Surgeon General's 1964 report on smoking and health \citep{USPHS}, smoking incidence in the US has fallen dramatically. CDC estimates that about 13.7\% of US adults, about 34.2 million, currently smoke cigarettes \citep{creamer}. While the number of smokers has fallen from about 40-45\% during the 1950s \citep{Saad}, tens of millions of Americans still smoke. Although more than half of all smokers make a quit attempt every year, unfortunately, smoking cessation success rates are quite low \citep{fiore}.

Currently, there are numerous pharmacotherapies and treatment strategies to aid current smokers in smoking cessation. Two of the most promising pharmacotherapies are varenicline and combination nicotine replacement therapy (cNRT) which combines the use of the nicotine patch and ad lib nicotine replacement such as nicotine gum or nicotine lozenges \citep{piper}. WSHS2 was the first comparative efficacy trial to compare the effects of the two most effective smoking cessation pharmacotherapies: varenicline versus cNRT with an active, less intense comparator treatment, the nicotine patch \citep{baker}. Baker et al. reported no significant differences in the effects of the less intense nicotine patch compared to varenicline and cNRT on smoking abstinence assessed using the self-reported and biochemically validated 7-day point-prevalence abstinence. 

The focus of Baker et al., was to determine the long-term abstinence effects of these three pharmacotherapies. However, it is also important to understand the various facets of the cessation process and how these might be influenced by different smoking cessation pharmacotherapies. WSHS2 utilized ecological momentary assessment  (EMA; \cite{shiffman}) to assess real-time pharmacotherapy use, withdrawal symptoms, cessation fatigue, and other constructs in the real-world setting during the initial cessation period, when relapse is most likely. Through analysis of EMA data, a deeper understanding of the cessation process might be achieved. For instance, potential triggers of relapse could be identified and in the future could be used to deliver a just-in-time adaptive intervention. Collection of EMA data has increased over the last decade, but the methodology to fully exploit the information contained in such data has not kept pace; thus the intensive longitudinal data provided by EMA are a rich resource that remains underutilized in understanding the dynamics of smoking relapse.

This research is a follow-up to that main outcome paper that will examine the rich EMA data to attempt to understand the lack of significant long-term outcome effects. Specifically, we will use the EMA data to examine how smoking behavior and other behaviors associated with smoking change as the subjects begin treatment and initiate their cessation attempt. Specifically, we want to explore how adherence to the assigned treatment regimen affects smoking cessation (outcome measured using daily self-reported cigarette counts) and whether this ``adherence'' effect is different for each treatment.  

Adherence is an important first step in our investigation because, presumably, intensive smoking cessation therapies, especially those that use a \textit{ad libitum} medication, are more difficult to comply with compared to less intensive therapies (i.e., putting on a patch in the morning). Thus, it could be that subjects are much more adherent to the use of nicotine patch compared to varenicline and  cNRT. Suppose the smoking cessation effect of varenicline or cNRT for subjects who comply to the treatment throughout the treatment period is different from subjects who do not comply at all. If this compliance effect is clinically much more pronounced compared to the compliance effect from nicotine patch use, then presumably, the lower likelihood of complying to varenicline and cNRT will reduce their exhibited treatment effect while the higher likelihood of compliance to the use of a nicotine patch will increase the treatment effect. This would result in a bias toward the null i.e. no difference in the effect of nicotine patch, varenicline, and cNRT, which was the conclusion reported in the original study \citep{baker}. Conversely, it could be that once a smoker relapses they no longer adhere to their treatment regimen. The use of EMA data will allow us to examine which effect occurs first. Hence, our objective is to explore whether there is a difference in the compliance pattern between the three treatments and more importantly, estimate the effect of compliance on smoking abstinence for each of the three different treatments.

To investigate the compliance effect, we used what we refer to as the average compliance effect (ACE) which is analogous to the usual average treatment effect (ATE) that compares the average difference in potential outcomes of the treatment versus control (or another different treatment). In our context, this will be the potential outcome under compliance to the treatment versus non-compliance to the treatment. Although our treatment was randomized, compliance is not and hence, is likely to be confounded. In addition, compliance may affect several variables which in turn may affect our outcome of interest,  self-reported number of cigarettes per day. In particular, cessation fatigue (i.e., the fatigue one experiences with the process of trying to quit smoking; \cite{piasecki}) has been shown to be an important predictor of relapse \citep{liu}.  Given that varenicline and cNRT are potentially more intensive therapies than the nicotine patch alone, compliance with the three pharmacotherapies may affect cessation fatigue. Thus, cessation fatigue may be an important mediator of the effect of compliance on smoking abstinence. In addition to a mediator, there may be potential time-varying confounders. For example, it is likely that compliance affects negative affect and craving which in turn affects cessation fatigue (i.e., the mediator) which then affects the outcome. In addition, it is also likely that negative affect and craving directly affect the number of cigarettes smoked per day. Thus, negative affect and craving are potential time-varying confounders of the mediator and outcome that have themselves been affected by compliance to the treatments. In summary, we are faced with the following four challenges when estimating the ACE: confounded time- varying exposure (compliance), time-varying exposure affected confounder of the mediator- outcome  relationship (i.e.,  negative  affect  and  craving),  time-varying  mediator (cessation fatigue), and time-varying outcome (cigarette  count) which itself implies time-dependent confounding. Figure \ref{dag} illustrates this using a simple 2 timepoint situation.

Presence of a confounded time-varying exposure and time-dependent confounding due to a time-varying outcome can usually be accounted for by using a marginal structural model (MSM) approach or the G-formula \citep{robins}. However, both approaches, in their original form, were not designed to handle the estimation of ACE in the presence of time-varying mediators and confounders of the mediator-outcome relationship which are themselves affected by the time-varying exposure. Here, we propose to solve these various issues and estimate the ACE using a missing data perspective. The approach of implementing an MSM using a missing data perspective was first highlighted by \cite{elliott} and later adopted for use in time-varying causal analysis by \cite{zhou} and \cite{tan}. \cite{elliott} described an alternative way to view Bayesian MSMs as a missing data problem but they did not formally introduce this idea for time-varying causal inference. \cite{zhou} formally extended the idea of \cite{elliott} to time-varying causal inference and proposed a doubly-robust method to ensure that the estimated ATE was valid. \cite{tan} employed similar ideas but focused on the truncation by death problem instead by imputing the missing potential principal strata. Although \cite{zhou} and \cite{tan} both considered time-varying causal inference problems, they did not address the presence of time-varying mediators or exposure affected confounders of the mediator-outcome relationship. 

In this work, we extend the ideas of \cite{elliott}, \cite{zhou}, and \cite{tan} by taking a missing data perspective. We handled time-varying mediators and exposure affected confounders of the mediator-outcome relationship by imputing their potential outcomes. We employ simulations to investigate some of the empirical statistical properties of our proposed approach and then apply this method to the EMA data of the WSHS2 study to investigate separately, the compliance effect of the three pharmacotherapies, nicotine patch, varenicline, and cNRT, on smoking cessation. We then conclude with a discussion of the implications and limitations of our results and findings.

\section{Method}
\subsection{Notation and setup}
Our proposed method is developed for the estimation of the effect of compliance for a single treatment.  Therefore, to estimate the compliance effect for nicotine patch, varenicline, and cNRT, we implement our proposed method on our data stratified by these three different treatment groups. Let $V$ be the baseline variables collected in our data set: age at consent, gender, race, education, length of longest quit time, cigarettes per day, The Fagerstrom Test for Nicotine Dependence (FTND) score \citep{heatherton}, and any history of mental health issues including: depression, bipolar disorder, schizophrenia, anxiety disorder, panic disorder, post-traumatic stress disorder, and attention deficit disorder. We denote $A(t)$ as whether the subject complied to the allocated treatment at time $t$ (i.e., our exposure of interest). Let $C_{1A}(t)$ and $C_{2A}(t)$ denote the potential outcomes for negative affect and craving respectively; the exposure affected confounders of the mediator-outcome relationship under exposure $A(t)$ at time $t$. For our mediator, cessation fatigue, we use $M_A(t)$ to denote its potential outcome at time $t$ under exposure $A(t)$. We denote the potential outcomes under exposure $A(t)$ at time $t$, as $Y_A(t)$. We use the $\bar{Z}=\{Z(1),\ldots,Z(T)\}$ notation to denote all the historical profile of a variable and $\bar{Z}(t)=\{Z(1),\ldots,Z(t)\}$ to denote the historical profile until time $t$. Hence,
$\bar{A}(t)=\{A(1),\ldots,A(t)\}$, $\bar{A}=\{A(1),\ldots,A(T)\}$ and $\bar{Y}_{\bar{A}}(t)=\{Y_A(1),\ldots,Y_A(t)\}$, $\bar{Y}_{\bar{A}}=\{Y_A(1),\ldots,Y_A(T)\}$. Notation of historical potential values for mediators and exposure affected confounders of the mediator-outcome relationship are similar. Finally, we assume that we observe these variables in the following order: $V$, $A(1)$, $C_{1A}(1)$, $C_{2A}(1)$, $M_A(1)$, $Y_A(1)$, $\ldots$, $A(t)$, $C_{1A}(t)$, $C_{2A}(t)$, $M_A(t)$, $Y_A(t)$. Figure \ref{dag} shows the resulting directed acyclic graph (DAG) produced for 2 timepoints ($T=2$). For $T\geq 2$, the DAG extends in a similar fashion. 

\begin{landscape}

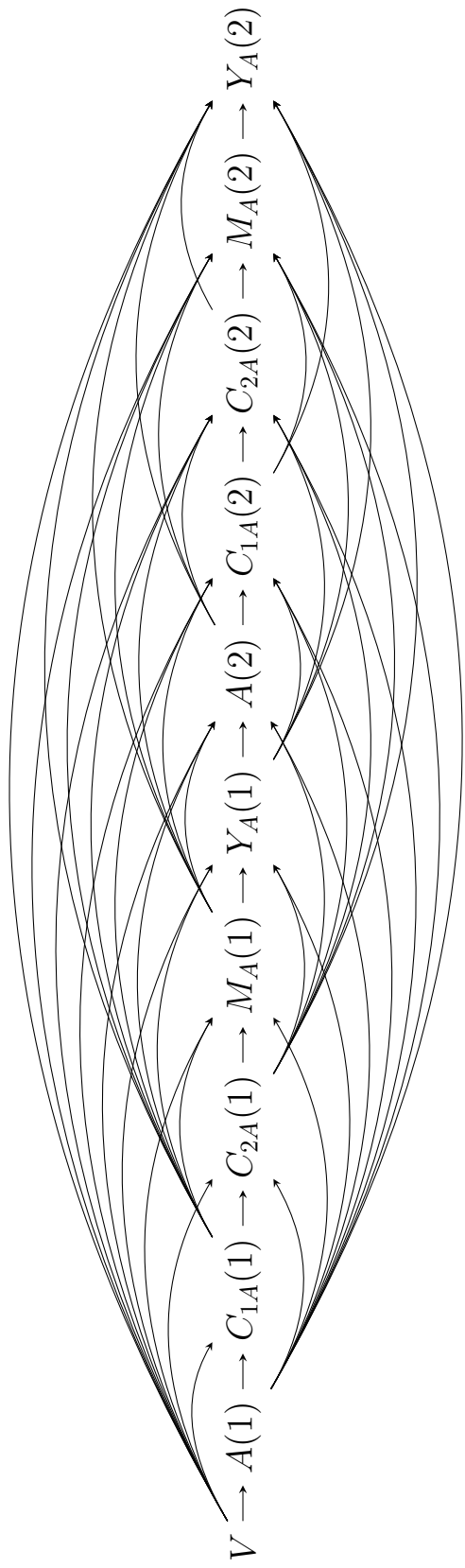
\begin{figure}[H]
	\caption{Directed acyclic graph for effect of compliance to a single treatment on daily cigarette count for two time points.\label{dag}}
	\large{\begin{tikzpicture}[%
		->,
		>=stealth,
		node distance=0.5cm,
		pil/.style={
			->,
			thick,
			shorten =1pt,}
		]
		\node (1) {$V$};
		\node[right=of 1] (2) {$A(1)$};
		\node[right=of 2] (3) {$C_{1A}(1)$};
		\node[right=of 3] (4) {$C_{2A}(1)$};
		\node[right=of 4] (5) {$M_A(1)$};
		\node[right=of 5] (6) {$Y_A(1)$};
		\node[right=of 6] (7) {$A(2)$};
		\node[right=of 7] (8) {$C_{1A}(2)$};
		\node[right=of 8] (9) {$C_{2A}(2)$};
		\node[right=of 9] (10) {$M_A(2)$};
		\node[right=of 10] (11) {$Y_A(2)$};
		
		\draw [->] (1.east) -- (2.west);
		\draw [->] (1) to [out=30, in=150] (3);
		\draw [->] (1) to [out=30, in=150] (4);
		\draw [->] (1) to [out=30, in=150] (5);
		\draw [->] (1) to [out=30, in=150] (6);
		\draw [->] (1) to [out=30, in=150] (7);
		\draw [->] (1) to [out=30, in=150] (8);
		\draw [->] (1) to [out=30, in=150] (9);
		\draw [->] (1) to [out=30, in=150] (10);
		\draw [->] (1) to [out=30, in=150] (11);
		
		\draw [->] (2.east) -- (3.west);
		\draw [->] (2) to [out=-30, in=-150] (4);
		\draw [->] (2) to [out=-30, in=-150] (5);
		\draw [->] (2) to [out=-30, in=-150] (6);
		\draw [->] (2) to [out=-30, in=-150] (7);
		\draw [->] (2) to [out=-30, in=-150] (8);
		\draw [->] (2) to [out=-30, in=-150] (9);
		\draw [->] (2) to [out=-30, in=-150] (10);
		\draw [->] (2) to [out=-30, in=-150] (11);
		
		\draw [->] (3.east) -- (4.west);
		\draw [->] (3) to [out=30, in=150] (5);
		\draw [->] (3) to [out=30, in=150] (6);
		\draw [->] (3) to [out=30, in=150] (7);
		\draw [->] (3) to [out=30, in=150] (8);
		\draw [->] (3) to [out=30, in=150] (9);
		\draw [->] (3) to [out=30, in=150] (10);
		\draw [->] (3) to [out=30, in=150] (11);
		
		\draw [->] (4.east) -- (5.west);
		\draw [->] (4) to [out=-30, in=-150] (6);
		\draw [->] (4) to [out=-30, in=-150] (7);
		\draw [->] (4) to [out=-30, in=-150] (8);
		\draw [->] (4) to [out=-30, in=-150] (9);
		\draw [->] (4) to [out=-30, in=-150] (10);
		\draw [->] (4) to [out=-30, in=-150] (11);
		
		\draw [->] (5.east) -- (6.west);
		\draw [->] (5) to [out=30, in=150] (7);
		\draw [->] (5) to [out=30, in=150] (8);
		\draw [->] (5) to [out=30, in=150] (9);
		\draw [->] (5) to [out=30, in=150] (10);
		\draw [->] (5) to [out=30, in=150] (11);
		
		\draw [->] (6.east) -- (7.west);
		\draw [->] (6) to [out=-30, in=-150] (8);
		\draw [->] (6) to [out=-30, in=-150] (9);
		\draw [->] (6) to [out=-30, in=-150] (10);
		\draw [->] (6) to [out=-30, in=-150] (11);
		
		\draw [->] (7.east) -- (8.west);
		\draw [->] (7) to [out=30, in=150] (9);
		\draw [->] (7) to [out=30, in=150] (10);
		\draw [->] (7) to [out=30, in=150] (11);
		
		\draw [->] (8.east) -- (9.west);
		\draw [->] (8) to [out=-30, in=-150] (10);
		\draw [->] (8) to [out=-30, in=-150] (11);
		
		\draw [->] (9.east) -- (10.west);
		\draw [->] (9) to [out=30, in=150] (11);
		
		\draw [->] (10.east) -- (11.west);
	\end{tikzpicture}}
\end{figure}
\end{landscape}	
	
\subsection{Missing data perspective}
We use a hypothetical data set to illustrate our motivation and rationale for using a missing data perspective to estimate the ACE. Consider two time points (i.e., $T=2$). Then, a hypothetical WSHS2 data set that we might have observed is given in Table \ref{miss_dat}.  In this table, `x' denotes values that are observed and `?' denotes values that are missing. Extension to $T>2$ is straightforward.

\begin{landscape}
\begin{table}[H]
	\caption{Table illustrating the missing data perspective in a longitudinal study with 2 time points, `x' -- observed; `?' --  missing. \label{miss_dat}}
	\footnotesize
	\begin{tabular}{|c|c|c|c|c|c|c|c|c|c|c|c|c|c|c|c|c|c|c|}
		\hline
		$V$ & $A(1)$ & $C_{11}(1)$ & $C_{10}(1)$ & $C_{21}(1)$ & $C_{20}(1)$ & $M_1(1)$ & $M_0(1)$ & $Y_1(1)$ & $Y_0(1)$ & $A(2)$ & $C_{11}(2)$ & $C_{10}(2)$ & $C_{21}(2)$ & $C_{20}(2)$ & $M_1(2)$ & $M_0(2)$ & $Y_1(2)$ & $Y_0(2)$ \\
		\hline
		x & 1 & x & ? & x & ? & x & ? & x & ? & 1 & x & ? & x & ? & x & ? & x & ? \\
		\hline
		x & 1 & x & ? & x & ? & x & ? & x & ? & 1 & x & ? & x & ? & x & ? & x & ? \\
		\hline
		x & 1 & x & ? & x & ? & x & ? & x & ? & 0 & ? & x & ? & x & ? & x & ? & x \\
		\hline
		x & 1 & x & ? & x & ? & x & ? & x & ? & 0 & ? & x & ? & x & ? & x & ? & x \\
		\hline
		x & 0 & ? & x & ? & x & ? & x & ? & x & 1 & x & ? & x & ? & x & ? & x & ? \\
		\hline
		x & 0 & ? & x & ? & x & ? & x & ? & x & 1 & x & ? & x & ? & x & ? & x & ? \\
		\hline
		x & 0 & ? & x & ? & x & ? & x & ? & x & 0 & ? & x & ? & x & ? & x & ? & x \\
		\hline
		x & 0 & ? & x & ? & x & ? & x & ? & x & 0 & ? & x & ? & x & ? & x & ? & x \\
		\hline
	\end{tabular}
\end{table}
\end{landscape}

Table \ref{miss_dat} suggests that we are faced with a data set that has 50\% of the data, excluding the observed baseline variables and the exposure, missing. Thus, a plausible strategy is to impute the counterfactuals. We refer to a counterfactual as the value that would have been observed if the subject, counter to the fact, was assigned a different treatment or exposure compared to the one we observed. In order to make this imputation valid, assumptions need to be made so that we will be able to use the observed values to construct a model to impute the missing values. Subsequently, we need to construct valid modeling algorithms or strategies such that the resulting ATE (ACE in our context) estimate will be consistent. Imputation uncertainty can then be captured by applying Rubin's rules for combining multiple imputations. This idea and strategy is described in detail in \cite{zhou}.

\subsection{PENCOMP approach of \cite{zhou}}
The Penalized Spline of Propensity Methods for Treatment Comparison (PENCOMP) is a doubly robust method to estimate the ATE (ACE in our context) at time $T$ for longitudinal studies. It assumes the presence of $V$, $A(t)$, and $Y_A(t)$ (i.e., no mediators and no exposure affected confounders of the mediator-outcome relationship). Let $v$, $a(t)$, and $y(t)$ denote the observed baseline variables, exposure, and outcome. Then PENCOMP requires the following three assumptions to impute the missing counterfactuals,
\begin{equation}
	\label{assume1}
	P(a(t)|\bar{y}(t-1),\bar{a}(t-1),v)>0,
\end{equation}
which states that the current exposure at time $t$ has positive probability of being observed given the past historical outcomes, exposures, and baseline variables. The second assumption is no interference which states that $Y_A(t)=Y_a(t)$ (i.e., the potential outcome of subject $i$ is not affected by which exposure subject $j$ is allocated to where $i\neq j$). The final assumption is no unmeasured confounding and sequential randomization which states that 
\begin{equation}
	\label{assume3}
	Y_A(t)\bot A(t)|\bar{Y}_{\bar{A}}(t-1),\bar{A}(t-1),V
\end{equation}
(i.e., given the past historical potential values, the allocation of the exposure can be considered balanced/randomized with respect to the potential outcome).

With these assumptions in place, PENCOMP then proceeds to estimate the ATE as follows . A bootstrap of the data set is drawn and for each bootstrap, the probability of the observed exposure is modeled using logistic regression. A logistic transformation is then performed on the predicted probability of exposure. This logistic transformed propensity score is then used in the penalized splines of propensity prediction (PSPP), a doubly robust imputation method developed for imputation of missing continuous variables under the missing at random (MAR) assumption \citep{zhang} by treating the counterfactual as the missingness that needs to be imputed. Once the counterfactuals are obtained, the algorithm moves on to the next time point with the same procedure of first obtaining the propensity score, logistic transform it, and then use the logistic transformed predicted propensity score in the PSPP model to impute the counterfactual at the next time point. Once all counterfactuals at all time points have been imputed, another bootstrap is performed and the above sequential procedure is repeated again. Once $B$ bootstraps have been obtained, Rubin's combining rules for multiple imputation can then be used to obtain the ATE estimate at each time point as well as the 95\% confidence interval. Details of the PENCOMP algorithm can be found in \cite{zhou} as well as Appendix A.

Note that the purpose of constructing the probability of observed exposure is for eventual use in PSPP. It is entirely possible to replace PSPP with different modeling strategies, (e.g., either simple ones such as linear regression or more complicated ones such as Bayesian additive regression trees \citep[BART;][]{chipman}) as long as the proposed modeling strategies are valid (i.e., consistent estimation of the regression parameters can hypothetically be achieved given assumptions). Each strategy has its own merits when imputing the counterfactuals. The choice of which strategy to use for imputation of the counterfactuals does not conflict with the main motivation of PENCOMP, which is to view the counterfactuals as a missing data problem and impute them.  

\subsection{Proposed method}
Our proposed method follows \cite{zhou} closely. Our contribution is to extend the missing data approach to include the imputation of the counterfactual mediator and exposure affected confounders of mediator-outcome relationship. In order to construct valid imputation algorithms that include the mediator and exposure affected confounders of mediator-outcome relationship, we require the following additional assumptions:
\begin{equation}
	\label{med1}
	Y_A(t)\bot M_A(t)|\bar{M}_{\bar{A}}(t-1),\bar{C}_{2\bar{A}}(t),\bar{C}_{1\bar{A}}(t),\bar{Y}_{\bar{A}}(t-1),\bar{A}(t),V
\end{equation}
which states that the potential outcome and mediator at time $t$ can be considered balanced/randomized given the past historical values;
\begin{equation}
	\label{med2}
	M_A(t) \bot A(t)|\bar{M}_{\bar{A}}(t-1),\bar{C}_{2\bar{A}}(t-1),\bar{C}_{1\bar{A}}(t-1),\bar{Y}_{\bar{A}}(t-1),\bar{A}(t-1),V
\end{equation}
i.e., the mediator and exposure at time $t$ can be considered balanced/randomized given the past historical values; similar assumptions will be needed for $C_{1A}(t)$ and $C_{2A}(t)$ as well; finally
\begin{align}
	\label{con1}
	&f(Y_{\bar{A}}(t)|M_{\bar{A}}(t)=m,M_{\bar{A}'}(t),\bar{C}_{2\bar{A}}(t),\bar{C}_{1\bar{A}}(t),\bar{Y}_{\bar{A}}(t-1),\bar{A}(t),V)\nonumber\\
	&\quad =f(Y_{\bar{A}'}(t)|M_{\bar{A}}(t),M_{\bar{A}'}(t)=m,\bar{C}_{2\bar{A}}(t),\bar{C}_{1\bar{A}}(t),\bar{Y}_{\bar{A}}(t-1),\bar{A}(t),V)
\end{align}
which states that the distribution of the potential outcomes is ultimately dependent on the actual realized value of the mediator and not the distribution of the mediator under $\bar{A}$ or $\bar{A}'$. Equations \ref{med1} and \ref{med2} allow us to impute the counterfactual mediator and exposure affected confounder of the mediator-outcome relationship. The last assumption, Equation (\ref{con1}) allows us to overcome the identifiability issue (See \cite{vanderweele} for more details) caused by the presence of an exposure affected confounder of the mediator-outcome relationship. With these additional assumptions in place, we construct our imputation algorithm to estimate the ACE at each time $t$ as follows:
\begin{enumerate}
	\item For $t=1$, construct the following regression models (e.g., linear regression, BART) using the observed outcomes: 
	\begin{itemize}
		\item $C_{1A}(1)|A(1),V$
		\item $C_{2A}(1)|C_{1A}(1),A(1),V$
		\item $M_A(1)|C_{2A}(1),C_{1A}(1),A(1),V$
		\item $Y_A(1)|M_A(1),C_{2A}(1),C_{1A}(1),A(1),V$
	\end{itemize}
	\item Similarly for $t>1$, construct the following regression models (e.g., linear regression, BART) using the observed outcomes: 
	\begin{itemize}
		\item $C_{1A}(t)|\bar{A}(t),\bar{Y}_{\bar{A}}(t-1),\bar{M}_{\bar{A}}(t-1),\bar{C}_{2\bar{A}}(t-1),\bar{C}_{1\bar{A}}(t-1),V$
		\item $C_{2A}(t)|\bar{C}_{1\bar{A}}(t),\bar{A}(t),\bar{Y}_{\bar{A}}(t-1),\bar{M}_{\bar{A}}(t-1),\bar{C}_{2\bar{A}}(t-1),V$
		\item $M_A(t)|\bar{C}_{2\bar{A}}(t),\bar{C}_{1\bar{A}}(t),\bar{A}(t),\bar{Y}_{\bar{A}}(t-1),\bar{M}_{\bar{A}}(t-1),V$
		\item $Y_A(t)|\bar{M}_{\bar{A}}(t),\bar{C}_{2\bar{A}}(t),\bar{C}_{1\bar{A}}(t),\bar{A}(t),\bar{Y}_{\bar{A}}(t-1),V$
	\end{itemize}
	\item Create 2 data sets $\{1,v\}$ and $\{0,v\}$.
	\item Use the model for $C_{1A}(1)$ in Step 1 with $\{1,v\}$ and $\{0,v\}$ to draw $C_{11}(1)$ and $C_{10}(1)$ respectively.
	\item Create 2 data sets $\{C_{11}(1),1,v\}$ and $\{C_{10}(1),0,v\}$.
	\item Use the model for $C_{2A}(1)$ in Step 1 with $\{C_{11}(1),1,v\}$ and $\{C_{10}(1),0,v\}$ to draw $C_{21}(1)$ and $C_{20}(1)$ respectively.
	\item Create 2 data sets $\{C_{21}(1),C_{11}(1),1,v\}$ and $\{C_{20}(1),C_{10}(1),0,v\}$.
	\item Use the model for $M_A(1)$ in Step 1 with $\{C_{21}(1),C_{11}(1),1,v\}$ and $\{C_{20}(1),C_{10}(1),0,v\}$ to draw $M_1(1)$ and $M_0(1)$ respectively.
	\item Create 2 data sets $\{M_1(1),C_{21}(1),C_{11}(1),1,v\}$ and $\{M_0(1),C_{20}(1),C_{10}(1),0,v\}$.
	\item Use the model for $Y_A(1)$ in Step 1 with $\{M_1(1),C_{21}(1),C_{11}(1),1,v\}$ and $\{M_0(1),C_{20}(1),C_{10}(1),0,v\}$ to draw $Y_1(1)$ and $Y_0(1)$ respectively.
	\item For $t>1$,
	\begin{itemize}
		\item Create 2 data sets $\{\bar{1},\bar{Y}_{\bar{1}}(t-1),\bar{M}_{\bar{1}}(t-1),\bar{C}_{2\bar{1}}(t-1),\bar{C}_{1\bar{1}}(t-1),v\}$ and $\{\bar{0},\bar{Y}_{\bar{0}}(t-1),\bar{M}_{\bar{0}}(t-1),\bar{C}_{2\bar{0}}(t-1),\bar{C}_{1\bar{0}}(t-1),v\}$.
		\item Use model for $C_{1A}(t)$ in Step 2 with $\{\bar{1},\bar{Y}_{\bar{1}}(t-1),\bar{M}_{\bar{1}}(t-1),\bar{C}_{2\bar{1}}(t-1),\bar{C}_{1\bar{1}}(t-1),v\}$ and $\{\bar{0},\bar{Y}_{\bar{0}}(t-1),\bar{M}_{\bar{0}}(t-1),\bar{C}_{2\bar{0}}(t-1),\bar{C}_{1\bar{0}}(t-1),v\}$ to draw $C_{1\bar{1}}(t)$ and $C_{1\bar{0}}(t)$ respectively.
		\item Create 2 data sets $\{\bar{C}_{1\bar{1}}(t),\bar{1},\bar{Y}_{\bar{1}}(t-1),\bar{M}_{\bar{1}}(t-1),\bar{C}_{2\bar{1}}(t-1),v\}$ and $\{\bar{C}_{1\bar{0}}(t),\bar{0},\bar{Y}_{\bar{0}}(t-1),\bar{M}_{\bar{0}}(t-1),\bar{C}_{2\bar{0}}(t-1),v\}$.
		\item Use model for $C_{2A}(t)$ in Step 2 with $\{\bar{C}_{1\bar{1}}(t),\bar{1},\bar{Y}_{\bar{1}}(t-1),\bar{M}_{\bar{1}}(t-1),\bar{C}_{2\bar{1}}(t-1),v\}$ and $\{\bar{C}_{1\bar{0}}(t),\bar{0},\bar{Y}_{\bar{0}}(t-1),\bar{M}_{\bar{0}}(t-1),\bar{C}_{2\bar{0}}(t-1),v\}$ to draw $C_{2\bar{1}}(t)$ and $C_{2\bar{0}}(t)$ respectively.
		\item Create 2 data sets $\{\bar{C}_{2\bar{1}}(t),\bar{C}_{1\bar{1}}(t),\bar{1},\bar{Y}_{\bar{1}}(t-1),\bar{M}_{\bar{1}}(t-1),v\}$ and $\{\bar{C}_{2\bar{0}}(t),\bar{C}_{1\bar{0}}(t),\bar{0},\bar{Y}_{\bar{0}}(t-1),\bar{M}_{\bar{0}}(t-1),v\}$.
		\item Use model for $M_A(t)$ in Step 2 with $\{\bar{C}_{2\bar{1}}(t),\bar{C}_{1\bar{1}}(t),\bar{1},\bar{Y}_{\bar{1}}(t-1),\bar{M}_{\bar{1}}(t-1),v\}$ and $\{\bar{C}_{2\bar{0}}(t),\bar{C}_{1\bar{0}}(t),\bar{0},\bar{Y}_{\bar{0}}(t-1),\bar{M}_{\bar{0}}(t-1),v\}$ to draw $M_{\bar{1}}(t)$ and $M_{\bar{0}}(t)$ respectively.
		\item Create 2 data sets $\{\bar{M}_{\bar{1}}(t),\bar{C}_{2\bar{1}}(t),\bar{C}_{1\bar{1}}(t),\bar{1},\bar{Y}_{\bar{1}}(t-1),v\}$ and $\{\bar{M}_{\bar{0}}(t),\bar{C}_{2\bar{0}}(t),\bar{C}_{1\bar{0}}(t),\bar{0},\bar{Y}_{\bar{0}}(t-1),v\}$.
		\item Use model for $Y_A(t)$ in Step 2 with $\{\bar{M}_{\bar{1}}(t),\bar{C}_{2\bar{1}}(t),\bar{C}_{1\bar{1}}(t),\bar{1},\bar{Y}_{\bar{1}}(t-1),v\}$ and $\{\bar{M}_{\bar{0}}(t),\bar{C}_{2\bar{0}}(t),\bar{C}_{1\bar{0}}(t),\bar{0},\bar{Y}_{\bar{0}}(t-1),v\}$ to draw $Y_{\bar{1}}(t)$ and $Y_{\bar{1}}(t)$ respectively.
	\end{itemize}
	\item For each $t$, calculate the exposure effect as $ATE(t)=E[Y_{\bar{1}}(t)-Y_{\bar{0}}(t)]$ ($ACE(t)$ in our context) which is estimated using $\frac{1}{n}\sum_{i=1}^nY_{i\bar{1}}(t)-Y_{i\bar{0}}(t)$.
	\item Similarly, estimate the empirical variance of the $ATE(t)$ (i.e., $ACE(t)$).
	\item Repeat Steps 1-13 $B$ times.
	\item Use Rubin's rules to obtain the standard error estimate and hence, the 95\% confidence intervals.
\end{enumerate} 

Although our proposed algorithm and PENCOMP are slightly different, the core ideas are the same which is, to use all the observed information in the data set up until time $t$ (depending on the temporal order of the variables) to construct the prediction model for the potential outcomes and then use these models to predict the potential outcomes at time $t$. These predicted potential outcomes would then be used to predict the potential outcomes at the subsequent time points and this whole process can be extended to time $T$. We did not include propensity scores in our imputation model for simplicity. Inclusion of propensity scores is straight forward. Finally, the PENCOMP approach looks at a more general ATE in the sense that it allows the comparison of various treatment profiles (e.g., $\{0,1,0\}$ versus $\{1,0,0\}$ for $T=3$). We are more interested in comparing the difference in the compliance effect if all subjects comply until time $t$ (i.e., $\bar{a}(t)=\{1,\ldots,1\}$) versus the situation when all subjects do not comply until time $t$ (i.e., $\bar{a}(t)=\{0,\ldots,0\}$). This does not mean that our algorithm is restricted to only the comparison of these two profiles. Profile comparisons like $\{0,1,0\}$ versus $\{1,0,0\}$ are also possible and extending our algorithm to handle such situations is simple.

\section{Simulation}
To investigate the statistical properties of our proposed method, we use 500 simulations to study the empirical bias, root mean squared error (RMSE), 95\% coverage, and the average 95\% confidence interval length (AIL) of the ACE produced using our algorithm. We considered two situations, (i) when the effect in our true models is linear and (ii) when the effect in our true models is non-linear and contains complicated interactions. For these two situations, we used sample size $n=1\,000$, $V\sim Unif(0,1)$ for our baseline variable, and 5 time points, $T=5$. For (i), our true model was given as follows:
\begin{align*}
	&P[A(1)|V]=expit(V)\\
	&C_A(1)|A(1),V\sim N(0.5A(1)+0.3V,0.9^2)\\
	&M_A(1)|C_A(1),A(1),V\sim N(0.2C_A(1)+0.2A(1)+0.9V,1.1^2)\\
	&Y_A(1)|M_A(1),C_A(1),A(1),V\sim N(0.5M_A(1)+0.4C_A(1)+0.3A(1)+0.5V,1)\\	
\end{align*}
For $t>1$,
\begin{align*}
	&P[A(t)=1|Y_A(t-1),M_A(t-1),C_A(t-1),A(t-1),V]\\ &\quad=expit(0.05Y_A(t-1)+0.1M_A(t-1)+0.1C_A(t-1)+A(t-1)+0.15V)\\
	&C_A(t)|A(t),Y_A(t-1),M_A(t-1),C_A(t-1),A(t-1),V \\
	&\quad\sim N(0.2A(t)+0.1Y_A(t-1)+0.1M_A(t-1)+C_A(t-1)+0.2A(t-1)+0.15V,1.5^2)\\
	&M_A(t)|C_A(t),A(t),Y_A(t-1),M_A(t-1),C_A(t-1),A(t-1),V \\
	&\quad\sim N(0.1C_A(t)+0.2A(t)+0.1Y_A(t-1)+M_A(t-1)+0.1C_A(t-1)+0.1A(t-1)+0.2V,0.75^2)\\
	&Y_A(t)|M_A(t),C_A(t),A(t),Y_A(t-1),M_A(t-1),C_A(t-1),A(t-1),V \\
	&\quad\sim N(0.05M_A(t)+0.05C_A(t)+0.1A(t)+Y_A(t-1)+0.1M_A(t-1)+0.1C_A(t-1)+0.1A(t-1)\\&\quad+0.1V,2^2)
\end{align*}
Two quick remarks regarding our data generating setup. First, we have only included one exposure affected confounder of the mediator-outcome relationship. Inclusion of another exposure affected confounder of the mediator-outcome relationship should not affect our simulation conclusions. Second, we employed the same regression coefficients for time $t>1$. This reduces the variability of our model across time. Another approach would have been to induce some hierarchical structure which is beyond the scope of this work. We shall elaborate on this point in Section \ref{disc}. For situation (ii), our true model is,
\begin{align*}
	&P[A(1)|V]=expit(V)\\
	&M_A(1)|A(1),V\sim N(0.5A(1)+0.3V,0.9^2) \\
	&Y_A(1)|M_A(1),A(1),V\sim N(0.35\sin(M_A(1)A(1))+0.4M_A(1)+0.3A(1)+0.5V,1)\\
\end{align*}
For $t>1$,
\begin{align*}
	&P[A(t)|Y_A(t-1),M_A(t-1),A(t-1),V]\\
	&\quad=expit[0.05Y_A(t-1)+0.1M_A(t-1)+A(t-1)+0.15V]\\
	&M_A(t)|A(t),Y_A(t-1),M_A(t-1),A(t-1),V \\
	&\quad\sim N(0.2A(t)+0.1Y_A(t-1)+M_A(t-1)+0.2A(t-1)+0.15V,1.5^2)\\
	&Y_A(t)|M_A(t),A(t),Y_A(t-1),M_A(t-1),A(t-1),V\\
	&\quad\sim N(0.35\sin(M_A(t)A(t))+0.05M_A(t)+0.1A(t)+Y_A(t-1)+0.1M_A(t-1)\\
	&\quad+0.1A(t-1)+0.1V,1)
\end{align*}

For both situations, we compared the use of multiple linear regression (MLR) versus Bayesian additive regression trees (BART) for Steps 1 and 2 in our proposed algorithm. We expect both methods to be unbiased and produce the approximate nominal coverage under situation (i) since the true generating models include only linear main effects without interaction terms. However, for situation (ii), in which the  true underlying model includes non-linear main and interaction effects,  we expect BART to remain unbiased and produce the approximate nominal coverage but we do not expect MLR to remain unbiased and attain nominal coverage.

\subsection{Results}
The result for situation (i) is given in Table \ref{res1}. From this table, we can see that the bias for our proposed algorithm under MLR and BART were rather similar with BART having slightly more variation evident from the larger RMSE, more than nominal coverage, and wider AIL. This larger than usual variation is expected as BART is a much more complicated model compared to MLR and may not be the most efficient when trying to predict simple models (linear main effects without interactions).

\begin{table}[H]
	\caption{Bias, root mean squared error (RMSE), 95\% coverage, and average 95\% confidence interval length (AIL) for situation (i) using our proposed algorithm with multiple linear regression (MLR) or Bayesian additive regression trees (BART).\label{res1}}
	\begin{tabular}{cccccccccc}
		\hline
		& & \multicolumn{4}{c}{MLR} & \multicolumn{4}{c}{BART} \\
		$t$ & True ACE & Bias & RMSE & Coverage & AIL & Bias & RMSE & Coverage & AIL \\
		\hline
		1 & 0.65 & 0.0003 &	0.091 & 0.928 & 0.304 & 0.0005 & 0.125 & 0.994 & 0.442 \\
		2 & 1.02 & -0.006 & 0.157 & 0.984 & 0.605 & 0.004 & 0.218 & 0.998 & 0.821 \\
		3 & 1.56 & -0.011 & 0.233 & 0.952 & 0.822 & 0.010 & 0.305 & 0.988 & 1.087 \\
		4 & 2.29 & -0.015 & 0.317 & 0.922 & 1.031 & 0.019 & 0.401 & 0.978 & 1.354 \\
		5 & 3.28 & -0.019 & 0.413 & 0.872 & 1.265 & 0.009 & 0.513 & 0.960 & 1.653 \\
		\hline
	\end{tabular}
\end{table}

For situation (ii), given in Table \ref{res2}, when the true generating model includes non-linear main and interaction effects, MLR produces consistently under estimated ACEs with larger RMSE and severely less than nominal coverage. In contrast, BART still produces low bias for the ACE estimate, lower RMSE, and close to nominal coverage.

\begin{table}[H]
	\caption{Bias, root mean squared error (RMSE), 95\% coverage, and average 95\% confidence interval length (AIL) for situation (ii) using our proposed algorithm with multiple linear regression (MLR) or Bayesian additive regression trees (BART).\label{res2}}
	\begin{tabular}{cccccccccc}
		\hline
		& & \multicolumn{4}{c}{Linear} & \multicolumn{4}{c}{BART} \\
		$t$ & True ACE & Bias & RMSE & Coverage & AIL & Bias & RMSE & Coverage & AIL \\
		\hline
		1 & 0.64 & -0.259 &	0.271 & 0.064 & 0.300 & -0.003 & 0.111 & 0.970 & 0.34 \\
		2 & 1.01 & -0.379 & 0.398 & 0.088 & 0.456 & -0.005 & 0.200 & 0.980 & 0.65 \\
		3 & 1.39 & -0.457 & 0.486 & 0.160 & 0.576 & 0.005 & 0.280 & 0.960 & 0.84 \\
		4 & 1.84 & -0.526 & 0.568 & 0.246 & 0.685 & 0.013 & 0.361 & 0.920 & 1.01 \\
		5 & 2.37 & -0.599 & 0.659 & 0.294 & 0.794 & -0.001 & 0.446 & 0.880 & 1.18 \\
		\hline
	\end{tabular}
\end{table}

\section{Compliance effect on cigarette counts: analysis of WSHS2 data}

\subsection{Study details}

WSHS2 was a comparative efficacy trial that compared two of the most effective smoking cessation therapies: varenicline versus cNRT with an active, less intense comparator treatment, nicotine patch \citep{baker}. Subjects (N=1086) were recruited from Madison and Milwaukee from May 2012 to November 2015 and randomly assigned to one of these three pharmacotherapies. Pharmacotherapy was provided for 12 weeks along with 6 sessions of smoking cessation counseling. During the EMA period, subjects completed 1 morning, 1 evening, and 1 random EMA prompt during the day, every day for 1 week prior to the target quit date (TQD) and 2 weeks post-quit and then every other day for another 2 weeks. This implies that in total, we could potentially obtain 84 repeated measurements per variable, per subject. The variables measured during this EMA period can be classified into the following domains: compliance to treatment, negative affect, craving, cessation fatigue, and cigarette count. Such intensive longitudinal data provide us with an opportunity to use this rich data set to explore the dynamics of smoking cessation.
   
In this analysis, our goal is to estimate the effect of treatment compliance during the first 4 weeks post TQD on the number of cigarettes smoked per day separately for each treatment group: patch, varenicline, and cNRT. Compliance for the patch group is defined as whether the subject used the nicotine patch that particular day while compliance for the varenicline group is defined as whether the subject took at least 1 varenicline pill that particular day. Compliance for cNRT is defined as whether the subject used the nicotine patch and at least 4 nicotine lozenges that particular day. 

Our simulation results suggest that when the true generating model is a simple linear regression model, using MLR or BART for Steps 1 and 2 of our algorithm will produce unbiased ACE estimates for a particular treatment (i.e., varencline, cNRT, or nicotine patch only). However, when the true generating model is more complicated, using MLR for Steps 1 and 2 will produce biased ACE estimates but using BART will still produce unbiased ACE estimates. For our data set, like most data sets, we do not know the true generating model. Hence, we may conservatively assume that the true generating model likely consists of complicated main and interaction effects. This suggests the use of BART for Steps 1 and 2 of our algorithm, which we will focus on presenting in this section. Results for using MLR can be found in the Appendix B.

Before implementing the proposed approach, we first needed to deal with the intermittently missing compliance variable in addition to the missing counterfactuals. Many subjects were intermittently missing the compliance variable for some of the days. Note the difference between missing potential outcomes (counterfactuals) versus  missing compliance indicators. The former is structurally missing while the latter is often unplanned. To distinguish, we refer to the former as counterfactual and to the latter as intermittently missing compliance indicators.  

\subsection{Intermittently missing compliance indicators}
\begin{figure}[H]
	\caption{Overall and by treatment group compliance profile during ecological momentary assessment period (7 days prior to 28 days post Target Quit Date) \label{comp_plot}.}
	\centering
	\includegraphics[scale=0.75]{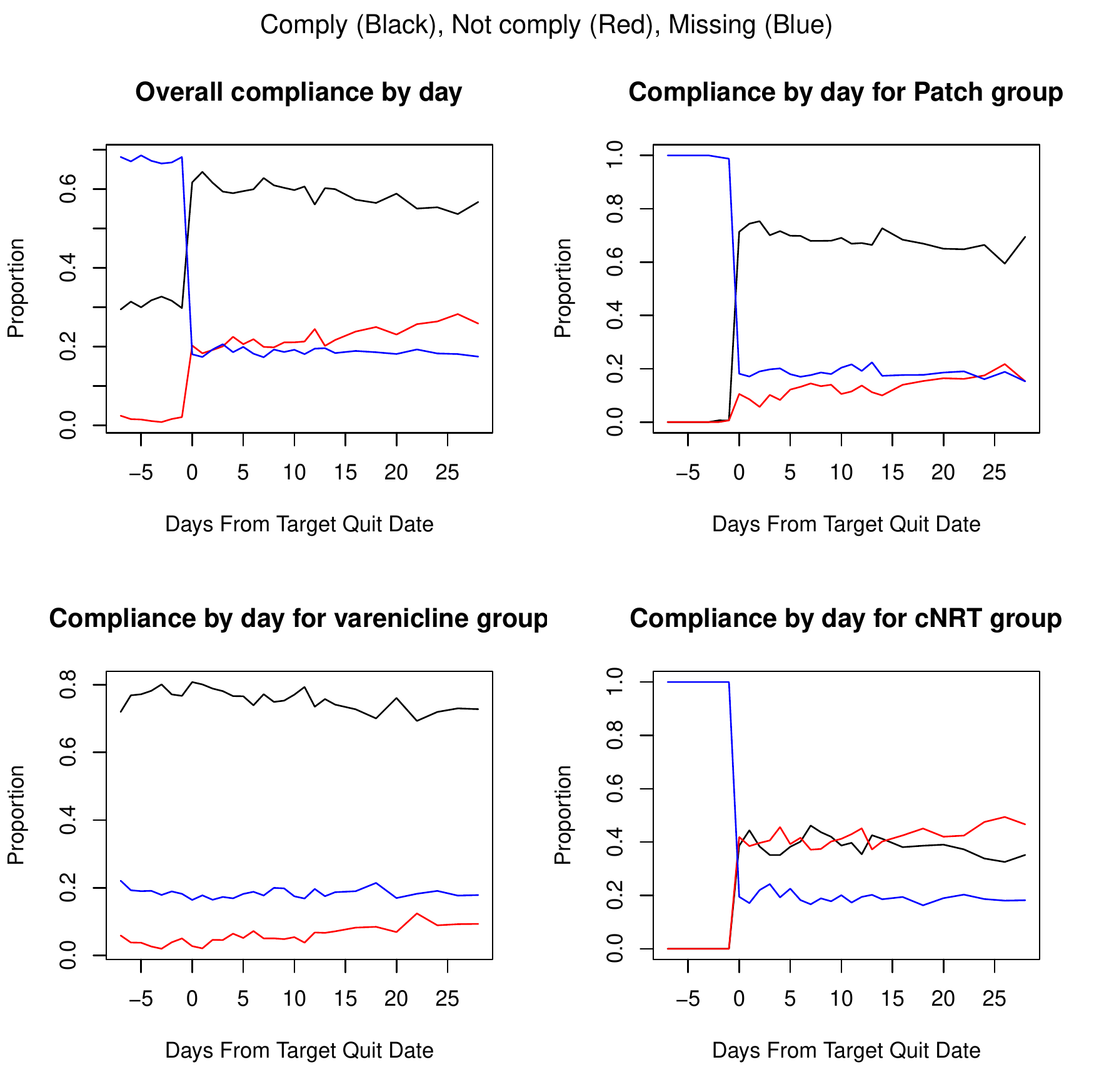}
\end{figure}

To get a general overall sense of the intermittent nature of our missing compliance indicators, we plotted the overall compliance profile throughout the EMA period as well as the same compliance profile stratified by each treatment. From Figure \ref{comp_plot}, we can see that prior to the TQD, most of the compliance indicators were missing. After the TQD, most subjects comply (about 60\% daily), some of the subjects do not comply (about 20\% daily), and some are missing (about 20\% daily). This suggests that we should focus our analysis on the data post TQD (i.e., 0 to 28 days after TQD).

%overall plus stratified by treatment plot

There are a number of reasons why subjects could be intermittently missing compliance indicators. For example, they could have missed the question or they might have forgotten whether they have complied to the treatment allocated for that day. Missingness due to these types of reasons can be considered missing completely at random and can be dealt with easily but more likely, the reason why compliance indicators are missing intermittently is due to factors tied to whether the subject complied to the treatment. For example, a subject may decide to not answer the question when they did not comply to the treatment in order to present themselves as `good' participants of the study. Such reasons suggest that compliance may be missing not at random (MNAR). 

To deal with potential MNAR missingness in our analysis, we first stratified our data set by the treatment each subject was allocated to: patch, varenicline, or cNRT. For each stratified group, we then performed the following sensitivity analysis: 1. imputing all intermittently missing compliance indicators as complied (all complied), 2. imputing all intermittently missing compliance indicators as not complied (all not complied), and finally 3. imputing all intermittently missing compliance indicators using Multiple Imputation by Chained Equations (MICE).  After imputing the compliance indicators, we then applied our proposed algorithm to estimate the ACE in each of the three different treatments and the effect on the daily cigarette count in the presence of a mediator and two exposure affected confounders of the mediator-outcome relationship.

In summary, we implemented our proposed method on our data set as follows. For each of the three different treatments: patch use, varenicline, cNRT, do the following:
\begin{enumerate}
	\item Impute the intermittently missing compliance indicators under the following three scenarios:
	\begin{enumerate}
		\item Complied,
		\item Not complied, or
		\item Using MICE.
	\end{enumerate}
	\item Using the data set with these imputed missing compliance indicators for each scenario, run Steps 1 and 2 of our algorithm using BART.
	\item Impute the potential outcomes for $C_{1A}(1)$, $C_{2A}(1)$, $M_A(1)$, and $Y_A(1)$ with $A(1)=0$ and $A(1)=1$ (Not complied versus complied) using Steps 3 to 10 of our algorithm.
	\item Given the potential outcomes at time $t=1$, use Step 11 to obtain the potential outcomes for $C_{1A}(2)$, $C_{2A}(2)$, $M_A(2)$, and $Y_A(2)$ under the compliance profile of $\bar{A}(2)=\{0,0\}$ and $\bar{A}(2)=\{1,1\}$ (i.e., never complied until time 2 versus always complied until time 2).
	\item Using a similar concept as in Step 4, extend the prediction of potential outcomes to $t=T$ (i.e., 28 days after TQD).
	\item For each $t$ until $T$, compute the empirical ACE.
	\item Repeat Steps 1-6 $B$ times and use Rubin's rules presented in Section 2.3 to compute the pointwise 95\% CI for the ACE estimate at each $t$.
\end{enumerate}

In addition, we investigated the difference in the potential daily cigarette count between varenicline group versus patch group (reference) as well as cNRT group versus patch group (reference) under the hypothetical situation where all subjects complied to their respective treatment as a further step to determine whether is it compliance or non-compliance that affects the treatment effect for each treatment group. The estimate for the potential cigarette count under each treatment can be obtained using the algorithm above while the 95\% confidence interval can be obtained using the 2.5 and 97.5 percentile of the bootstrapped values.  

\subsection{Results}
\begin{figure}[H]
	\caption{Compliance effect (never compliers versus compliers) for subjects allocated to nicotine patch. Missing compliance indicators were either imputed as complied (all compliers), not complied (All not compliers) or imputed using multiple imputation by chained equations (MICE; imputed compliers). Bayesian additive regression trees was used to impute the counterfactual. Dotted lines indicate pointwise 95\% confidence interval. \label{ptreat_bart}}
	\centering
	\includegraphics[scale=0.75]{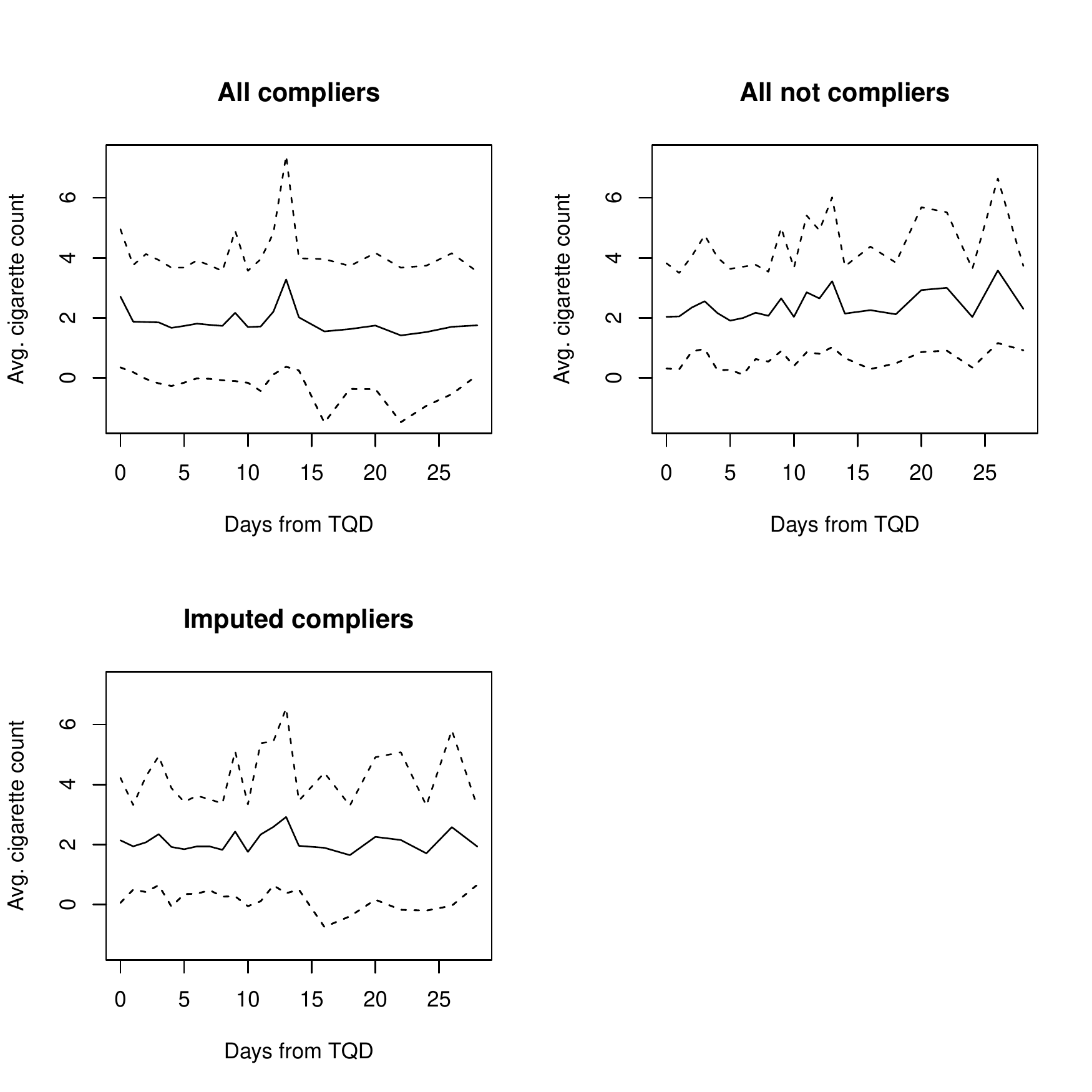}
\end{figure}

\begin{figure}[H]
	\caption{Compliance effect (never compliers versus compliers) for subjects allocated to varenicline. Missing compliance indicators were either imputed as complied (all compliers), not complied (All not compliers) or imputed using multiple imputation by chained equations (MICE; imputed compliers). Bayesian additive regression trees was used to impute the counterfactual. Dotted lines indicate pointwise 95\% confidence interval. \label{vtreat_bart}}
	\centering
	\includegraphics[scale=.75]{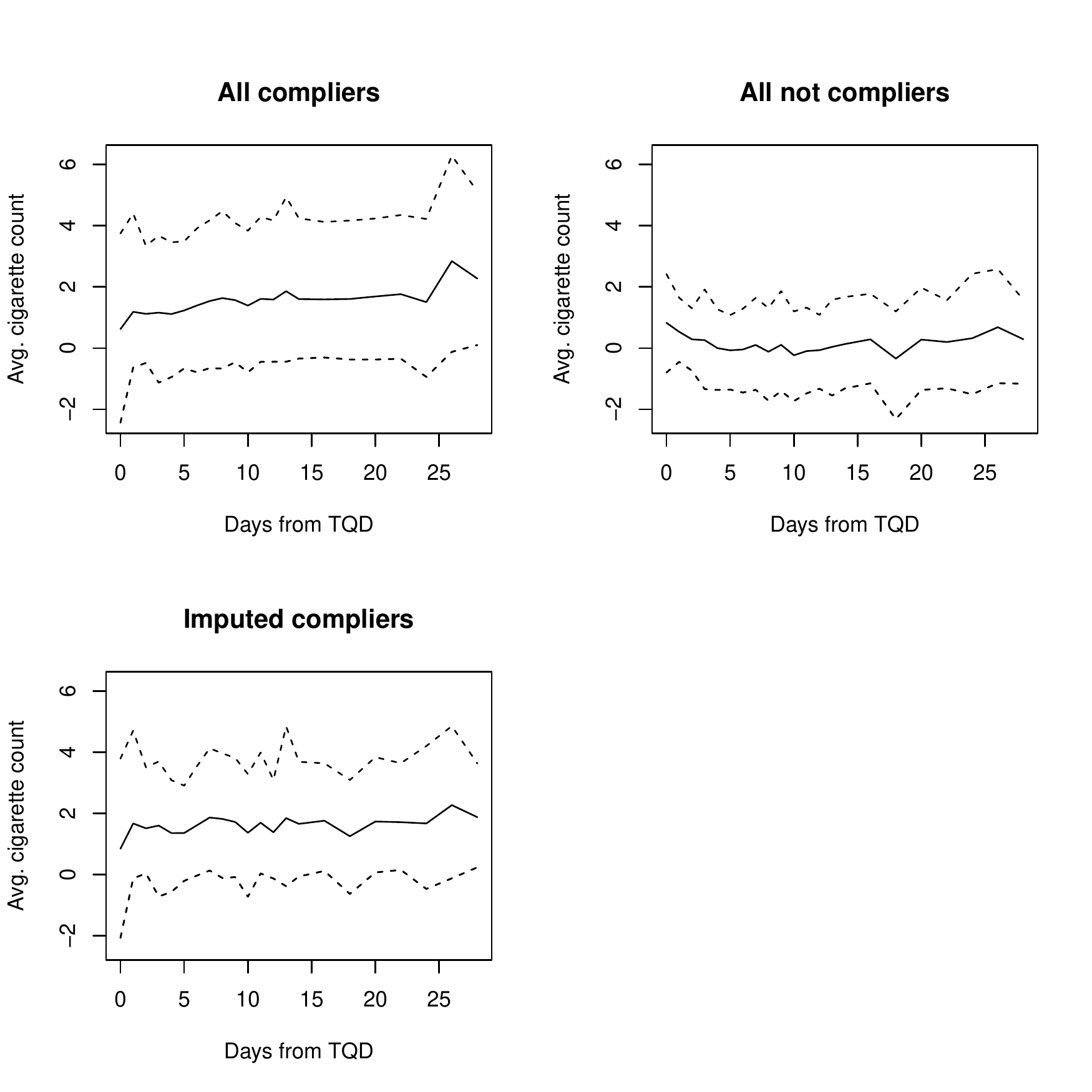}
\end{figure}

\begin{figure}[H]
	\caption{Compliance effect (never compliers versus compliers) for subjects allocated to cNRT. Missing compliance indicators were either imputed as complied (all compliers), not complied (All not compliers) or imputed using multiple imputation by chained equations (MICE; imputed compliers). Bayesian additive regression trees was used to impute the counterfactual. Dotted lines indicate pointwise 95\% confidence interval. \label{ctreat_bart}}
	\centering
	\includegraphics[scale=.75]{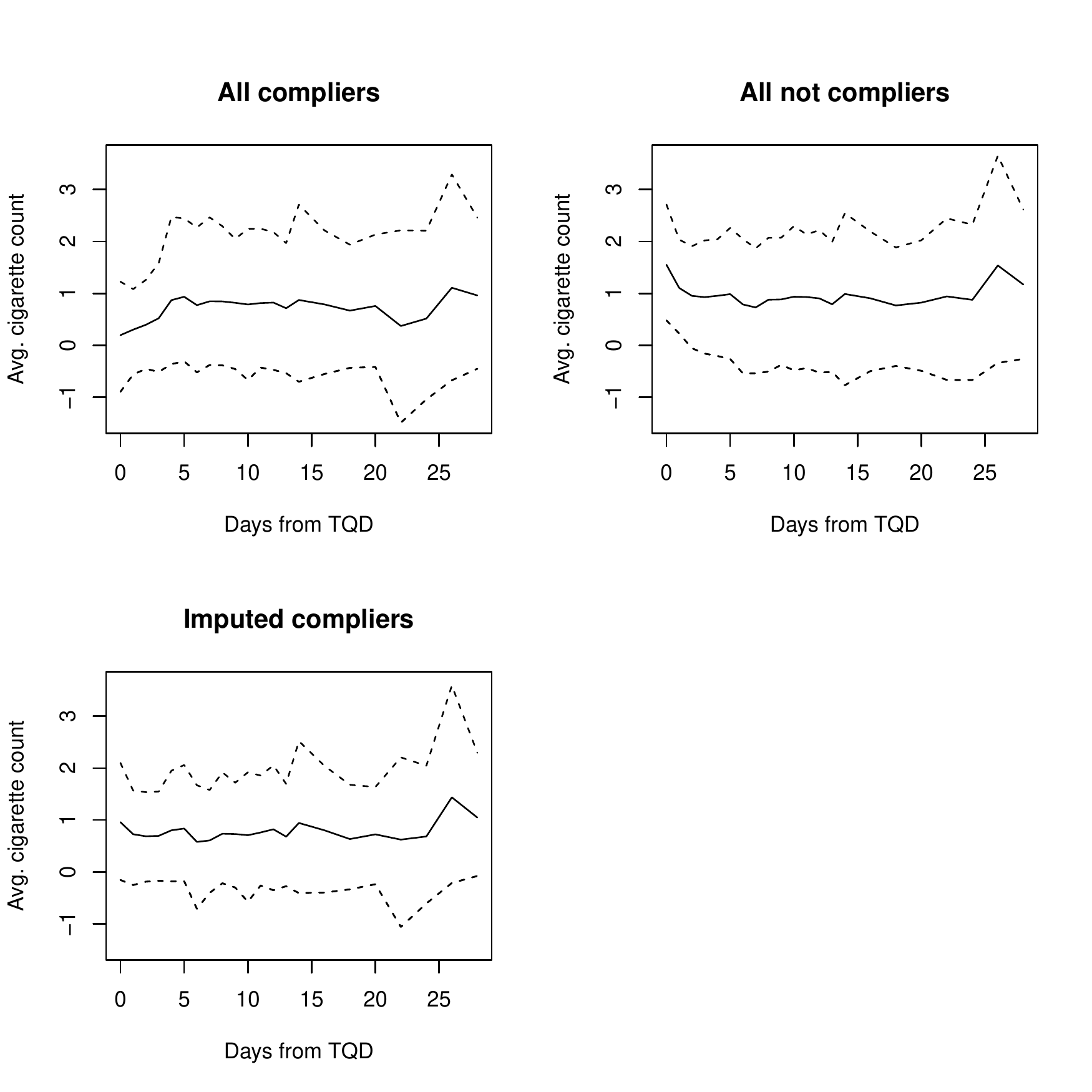}
\end{figure}

Figures \ref{ptreat_bart} to \ref{ctreat_bart} show the compliance effect on cigarette count for subjects allocated to patch, varenicline, and cNRT treatment respectively. The results suggest a slightly significant compliance effect for subjects allocated to the nicotine patch group regardless of how we imputed the missing compliance indicators. The estimated compliance effect was about 2 cigarettes per day on average, with some fluctuation. For varenicline, we see a rather steady and slightly variable profile fluctuating between 0.5 to 1.5 cigarettes per day. Finally, for cNRT we again see slight fluctuations of the compliance profile between 0 and 1.5 cigarettes per day. 

\begin{figure}[H]
	\caption{Difference in potential cigarette counts of varenicline versus Patch (Red) and cNRT versus Patch (Blue) with the Patch group as reference under the hypothetical situation where all subjects complied to treatment. Missing compliance indicators were imputed as complied, not complied, or using Multiple Imputation by Chained Equations (MICE). Dotted lines indicate pointwise 95\% confidence interval. \label{com_bart}}
	\centering
	\includegraphics[scale=.75]{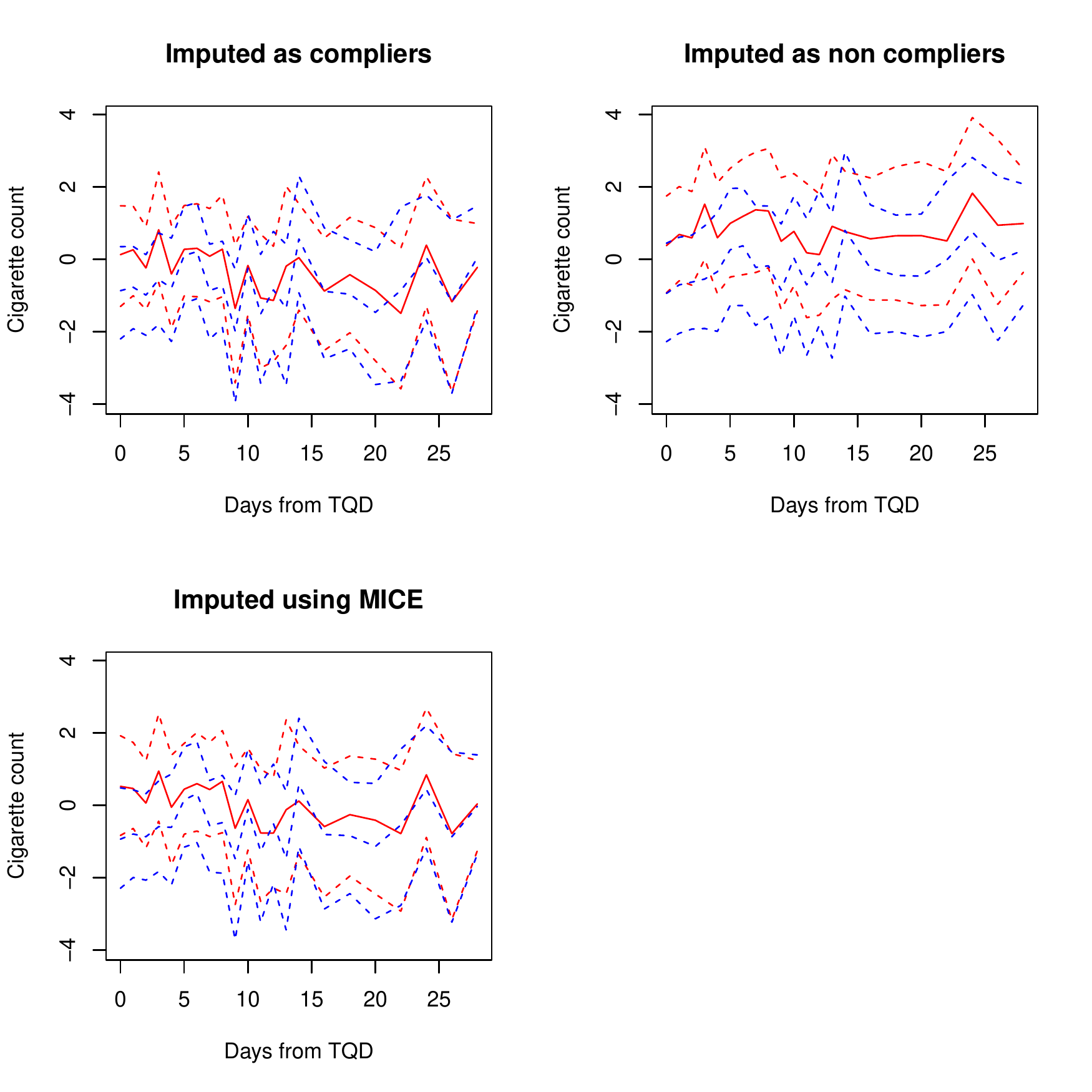}
\end{figure}

Figure \ref{com_bart} shows the difference in potential cigarette counts of varenicline versus Patch and cNRT versus Patch with the Patch group as reference under the hypothetical situation where all subjects complied to treatment. We can see that the difference for varenicline versus Patch and cNRT versus Patch is not significantly different from 0 and there is little separation between these two profiles throughout the 4 weeks post-quit. This suggests that under the hypothetical situation where all subjects complied to their respective treatments, there would be no significant difference between the potential cigarette count regardless of which smoking cessation treatment the subject is allocated to. 

\section{Discussion}
\label{disc}
WSHS2 was designed to compare both the short-term and long-term efficacy of two effective smoking treatments, varenicline and cNRT, versus an active comparator, nicotine patch. The original paper focused on the long-term efficacy of these three treatments and concluded that there were no significant differences between them in their effect on smoking abstinence. We investigated this conclusion further by considering how adherence to the treatment affected smoking abstinence through the use of rich EMA data collected during the first 4 weeks of the quit attempts. We observed that compliance for subjects allocated to the nicotine patch group and varenicline group were rather high, about 70 to 75\% (See Figure \ref{comp_plot}).  Compliance to cNRT in contrast, was lower at around 40\%. We found that only smokers allocated to the nicotine patch treatment had a significant compliance effect after taking into account the presence of mediators and exposure affected confounders  of the mediator-outcome relationship. These results suggest that compliance is likely not the reason for the treatment effect similarity of nicotine patch compared to varenicline and cNRT. It also suggests that we might want to explore how the mediator and exposure affected confounders of the mediator-outcome relationship affects the pathway between the three different treatments and smoking abstinence measured in terms of daily cigarette counts.

In our analysis, the outcome of interest was cigarette counts which is more naturally modeled using a zero-inflated Poisson (ZIP) regression (because of the large number, about 60\%, of 0 observed cigarette counts) rather than MLR or BART. We separately ran a small simulation to investigate how our proposed algorithm would perform if we replaced the model for Steps 2 and 3 of our algorithm with a ZIP regression. We found that the ACE estimates were still unbiased using simulation. We then proceeded to implement our proposed algorithm using ZIP regression on our data set. We obtained very similar ACE estimates as the ones produced by MLR. Hence, we decided that using a normal distribution for the analysis was sufficient and did not feel that count based models were needed for our data set.   

Another direction of plausible interest is to use hierarchical Bayesian methods to induce the ``smoothing'' of the coefficients over the 28 day EMA period. Currently, in order to induce smoothing, we forced our model to compute the same regression coefficient for $t>1$. This may have been a little too over-restrictive as the very smooth profiles in Figures \ref{ptreat} to \ref{ctreat} suggest. One possibility may be to use a Bayesian hierarchical model by assuming that each regression coefficient at time $t$ comes from the same underlying distribution. This idea can be extended to BART where perhaps ``similar'' tree structures will be obtained at each $t$. Finally, to reduce the reliance on the AR(1) assumption to help us with the collinearity issues, we could employ ``variable selection'' methods to decide which covariates to include as ``historical'' variables when modeling the exposure affected confounders of the mediator-outcome relationship, mediator, and outcome. We aim to address these issues in future work.

\bibliography{references}
\bibliographystyle{biom}

\appendix
\section{Detailed algorithm of the multiple imputation implementation of the Penalized Spline of Propensity Methods for Treatment Comparison (PENCOMP) algorithm}

This section of the appendix provides further details for the multiple imputation implementation of the Penalized Spline of Propensity Methods for Treatment Comparison (PENCOMP) algorithm proposed by \cite{zhou}.

\begin{enumerate}
	\item Draw a bootstrap of the data set.
	\item For each bootstrap, at time $t$, estimate  the probability of the observed exposure, i.e. $A(t)=a(t)$, with $a(t)$ as the binary outcome, $v$, $\bar{a}(t-1)$, and, $\bar{y}(t-1)$ as the covariates. Define $\hat{P}_{\bar{a}(t)}=\prod_{i=1}^t \hat{P}_{a(i)}$ and take the logistic transformation of $\hat{P}_{\bar{a}(t)}$.
	\item Together with the logistic transformation of $\hat{P}_{\bar{a}(t)}$, $v$, $\bar{a}(t)$, and, $\bar{y}(t-1)$ with $y(t)$ as the outcome, estimate a model for $E[Y(t)|\bar{y}(t-1),\bar{a}(t),v]$ using penalized splines of propensity prediction (PSPP), a doubly robust imputation method developed for imputation of missing continuous variables under the missing at random (MAR) assumption \citep{zhang}.
	\item Now for each $t$ with the corresponding treatment profiles of interest, impute the counterfactual (missing) exposure profile probabilities as well as the counterfactual outcomes.
	\item Let $\hat{\Delta}_{\bar{A}(T),\bar{A}'(T)}^{(b)}=E[Y_{\bar{A}}(T)-Y_{Y_{\bar{A}'}(T)}(T)]$, where $\bar{A}(T)\neq \bar{A}'$, with associated pooled variance estimates $W_{\bar{A}(T),\bar{A}'(T)}^{(b)}$, for each bootstrap $b$.
	\item The ATE estimate is then $\bar{\Delta}_{\bar{A}(T),\bar{A}'(T),B}=\sum_{b=1}^BB^{-1}\hat{\Delta}_{\bar{A}(T),\bar{A}'(T)}^{(b)}$, and the estimate of the variance for $\bar{\Delta}_{\bar{A}(T),\bar{A}'(T),B}$ is $T_B=\bar{W}_{\bar{A}(T),\bar{A}'(T),B}+(1+1/B)D_{\bar{A}(T),\bar{A}'(T),B}$, where $\bar{W}_{\bar{A}(T),\bar{A}'(T),B}=\sum_{b=1}^BB^{-1}W_{\bar{A}(T),\bar{A}'(T)}^{(b)}$, $D_{\bar{A}(T),\bar{A}'(T),B}=\sum_{b=1}^B\frac{(\hat{\Delta}_{\bar{A}(T),\bar{A}'(T)}^{(b)}-\bar{\Delta}_{\bar{A}(T),\bar{A}'(T),B})^2}{B-1}$. The ATE estimate then follows a $t$ distribution with degree of freedom $\nu$, i.e. $\frac{ATE-\bar{\Delta}_{\bar{A}(T),\bar{A}'(T),B}}{\sqrt{T_B}}\sim t_{\nu}$, where $\nu=(B-1)(1+\frac{\bar{W}_{\bar{A}(T),\bar{A}'(T),B}}{D_{\bar{A}(T),\bar{A}'(T),B}(B+1)})^2$.
\end{enumerate}

\section{Compliance analysis results using multiple linear regression as the imputation model}

\begin{figure}[H]
	\caption{Compliance effect (never compliers versus compliers) for subjects allocated to nicotine patch. Missing compliance indicators were either imputed as complied (all compliers), not complied (All not compliers) or imputed using multiple imputation by chained equations (MICE; imputed compliers). Multiple linear regression was used to impute the counterfactual. \label{ptreat}}
	\centering
	\includegraphics[scale=.75]{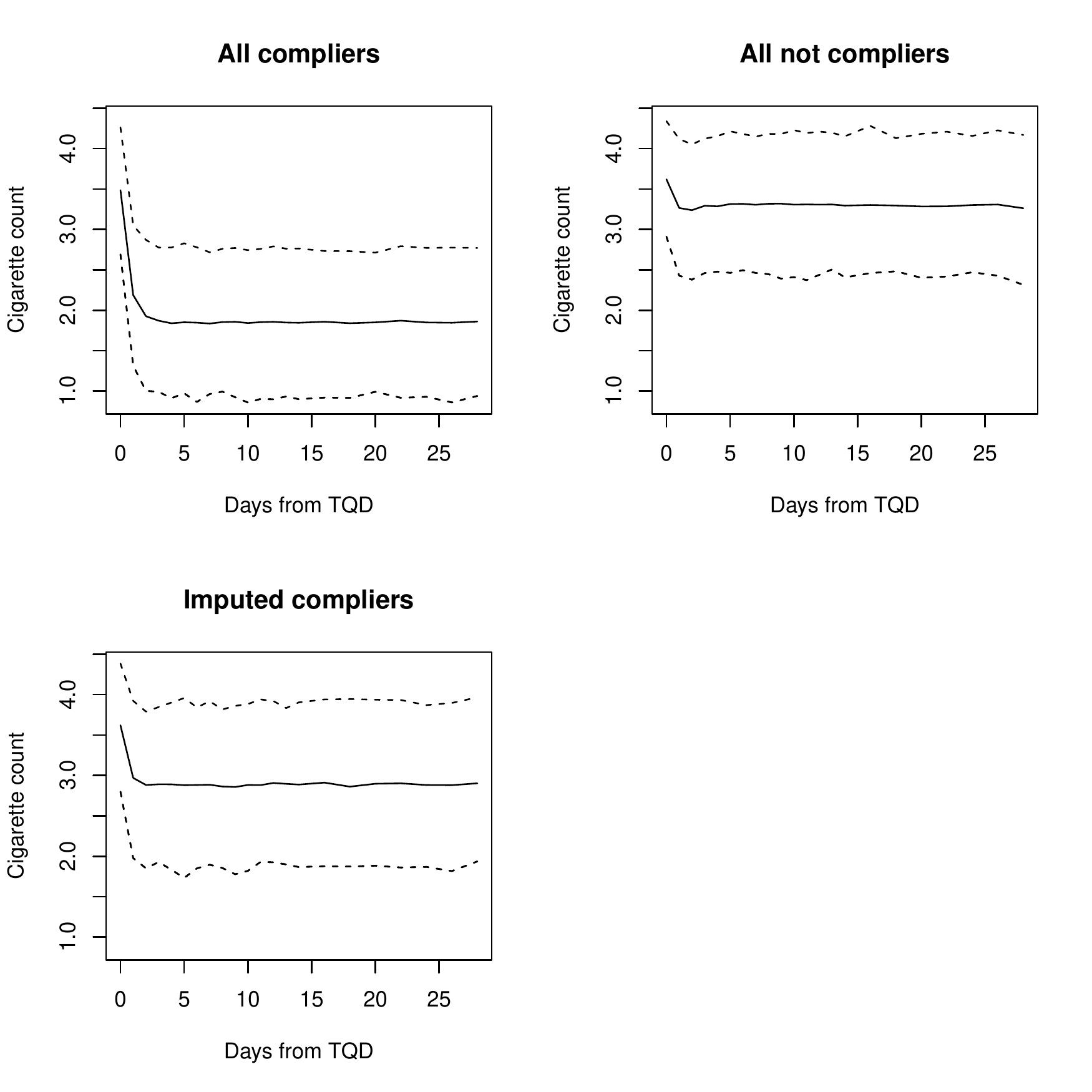}
\end{figure}

\begin{figure}[H]
	\caption{Compliance effect (never compliers versus compliers) for subjects allocated to varenicline. Missing compliance indicators were either imputed as complied (all compliers), not complied (All not compliers) or imputed using multiple imputation by chained equations (MICE; imputed compliers). Multiple linear regression was used to impute the counterfactual.\label{vtreat}}
	\centering
	\includegraphics[scale=.75]{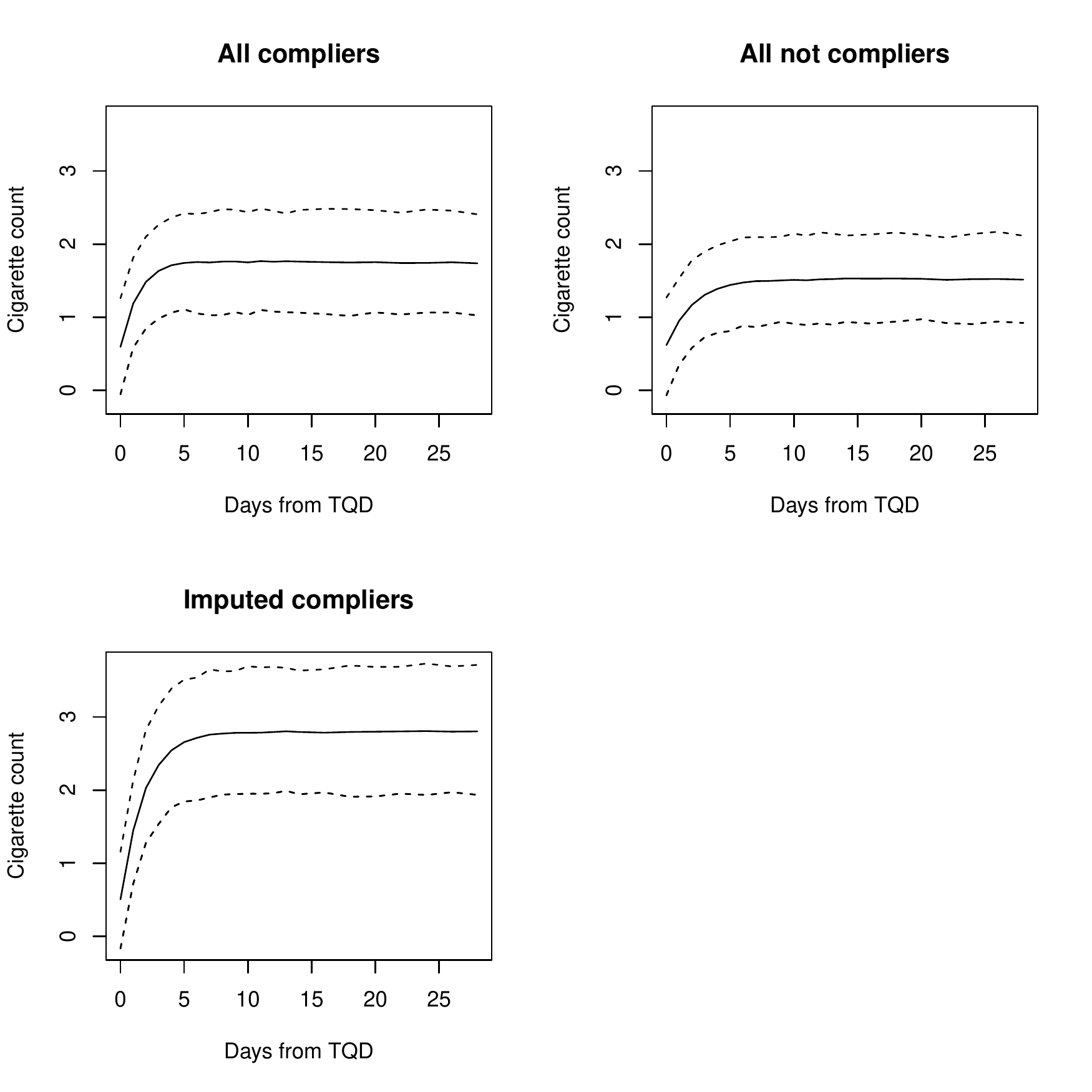}
\end{figure}

\begin{figure}[H]
	\caption{Compliance effect (never compliers versus compliers) for subjects allocated to cNRT. Missing compliance indicators were either imputed as complied (all compliers), not complied (All not compliers) or imputed using multiple imputation by chained equations (MICE; imputed compliers). Multiple linear regression used was to impute the counterfactual.\label{ctreat}}
	\centering
	\includegraphics[scale=.75]{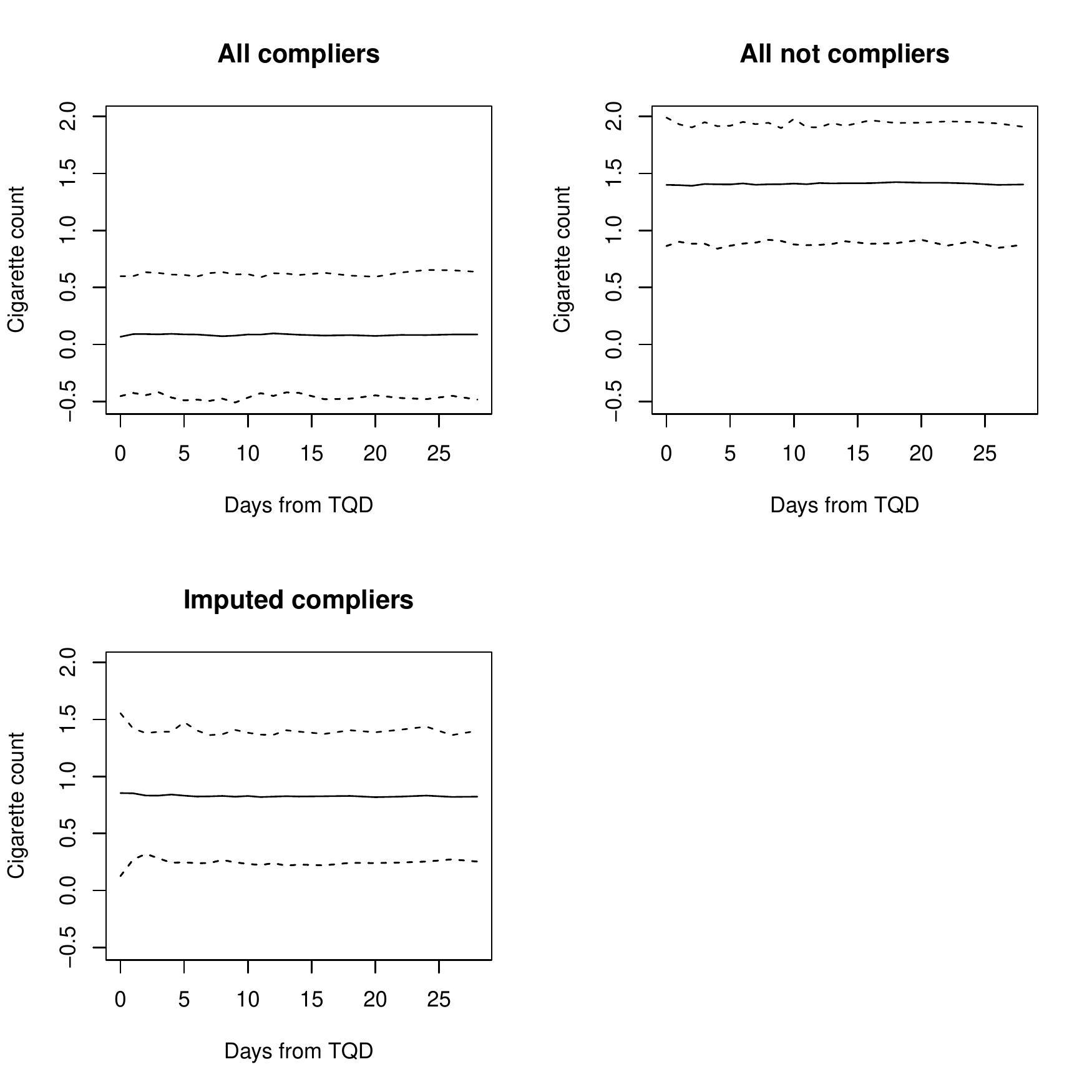}
\end{figure}
\end{document}